\documentclass[adp,fleqn]{w-art}
\usepackage{times}
\usepackage{w-thm}
\usepackage[]{graphicx}
\begin{document}

\renewcommand{\copyrightyear}{2005}

\def\lsim{\hbox{ \raise.35ex\rlap{$<$}\lower.6ex\hbox{$\sim$}\ }}
\def\gsim{\hbox{ \raise.35ex\rlap{$>$}\lower.6ex\hbox{$\sim$}\ }}

\def\xrightarrow#1#2#3#4{\,\lower#1pt\hbox{$\stackrel{\stackrel{\displaystyle #2}%
{\hbox to #3cm{\rightarrowfill}}}{#4}$}\,}

\DOIsuffix{theDOIsuffix}


\pagespan{1}{}

\Receiveddate{26 October 2005}


\title{The Revival of Cosmic Strings}

\author[Mairi Sakellariadou]{Mairi
     Sakellariadou\footnote{Corresponding author: e-mail: {\sf
     mairi.sakellariadou@kcl.ac.uk}, Phone: +44(0)2078481535 Fax:
     +44(0)2078482420}\inst{1}} \address[\inst{1}]{Department of
     Physics, King's College London, Strand, London WC2R 2LS, U.K. }

\begin{abstract}
Cosmic strings are one-dimensional topological defects which could
have been formed in the early stages of our Universe. They triggered a
lot of interest, mainly for their cosmological implications: they
could offer an alternative to inflation for the generation of density
perturbations. It was shown however that cosmic strings lead to
inconsistencies with the measurements of the cosmic microwave
background temperature anisotropies. The picture is changed
recently. It was shown that, on the one hand, cosmic strings can be
generically formed in the framework of supersymmetric grand unified
theories and that, on the other hand, cosmic superstrings could play
the r\^ole of cosmic strings.  There is also some possible
observational support. All this lead to a revival of cosmic strings
research and this is the topic of my lecture.
\end{abstract}

\maketitle                   
\section{Introduction}

Cosmic strings attracted a lot of interest around the eighties and
nineties. They offered an alternative mechanism to cosmological
inflation for the generation of the primordial density perturbations
leading to the large-scale structure formation one observes.  However,
towards the turn of the century cosmic strings lost their appeal,
since it was shown that they lead to inconsistencies with the Cosmic
Microwave Background (CMB) measurements. Nevertheless, the story of
cosmic strings does not end here. In the last few years there has been
a remarkable revival of the theoretical and observational activity.

In this lecture, I will discuss the present view on the cosmological
r\^ole of cosmic strings.  In Section 2, I will discuss aspects of
cosmic strings in the framework of Grand Unified Theories (GUTs).  I
will first analyse the formation and classification of topological as
well as embedded defects. I will then briefly discuss the CMB
temperature anisotropies and I will compare the predictions of
topological defects models with current measurements. I will then
conclude that topological defects in general, and cosmic strings in
particular, are ruled out as the unique source of density
perturbations leading to the observed structure formation. At this
point I do not conclude that cosmic strings are ruled out, but I ask
instead which are the implications for the models of high energy
physics which we employed to construct our cosmological scenario. The
first question is whether cosmic strings are expected to be
generically formed. I will address this question in the framework of
Supersymmetric Grand Unified Theories (SUSY GUTs). I will show that
cosmic strings are indeed expected to be generically formed within a
large class of models within SUSY GUTs and therefore one has to use
mixed models, consisting in inflation with a sub-dominant partner of
cosmic strings. I will then examine whether such mixed models are
indeed compatible with the CMB data. I will present two well-studied
inflationary models within supersymmetric theories, namely F/D-term
hybrid inflation. I will impose constraints on the free parameters of
the models (masses and couplings) so that there is agreement between
theory and measurements.  In Section 3, I will address the issue of
cosmic superstrings as cosmic strings candidates, in the context of
braneworld cosmologies.  In Section 4, I will discuss a candidate of a
gravitational lensing by a cosmic string.  I will round up with the
conclusions in Section 5.

\section{Topological Defects}

\subsection{Topological Defects in GUTs}

The Universe has steadily cooled down since the Planck time, leading
to a series of Spontaneously Broken Symmetries (SSB). SSB may lead to
the creation of topological defects \cite{td1,td2}, which are false
vacuum remnants, such as domain walls, cosmic strings, monopoles, or
textures, via the Kibble mechanism \cite{kibble}.

The formation or not of topological defects during phase transitions,
followed by SSB, and the determination of the type of the defects,
depend on the topology of the vacuum manifold ${\cal M}_n$.  The
properties of ${\cal M}_n$ are usually described by the $k^{\rm th}$
homotopy group $\pi_k({\cal M}_n)$, which classifies distinct mappings
from the $k$-dimensional sphere $S^k$ into the manifold ${\cal
M}_n$. To illustrate that, let me consider the symmetry breaking of a
group G down to a subgroup H of G . If ${\cal M}_n={\rm G}/{\rm H}$
has disconnected components, or equivalently if the order $k$ of the
nontrivial homotopy group is $k=0$, then two-dimensional defects,
called {\sl domain walls}, form.  The spacetime dimension $d$ of the
defects is given in terms of the order of the nontrivial homotopy
group by $d=4-1-k$. If ${\cal M}_n$ is not simply connected, in other
words if ${\cal M}_n$ contains loops which cannot be continuously
shrunk into a point, then {\sl cosmic strings} form. A necessary, but
not sufficient, condition for the existence of stable strings is that
the first homotopy group (the fundamental group) $\pi_1({\cal M}_n)$
of ${\cal M}_n$, is nontrivial, or multiply connected. Cosmic strings
are line-like defects, $d=2$. If ${\cal M}_n$ contains unshrinkable
surfaces, then {\sl monopoles} form, for which $k=1, ~d=1$.  If ${\cal
M}_n$ contains noncontractible three-spheres, then event-like defects,
{\sl textures}, form for which $k=3, ~d=0$.

Depending on whether the symmetry is local (gauged) or global (rigid),
topological defects are called local or global. The energy of local
defects is strongly confined, while the gradient energy of global
defects is spread out over the causal horizon at defect formation.
Patterns of symmetry breaking which lead to the formation of local
monopoles or local domain walls are ruled out, since they should soon
dominate the energy density of the Universe and close it, unless an
inflationary era took place after their formation.  Local textures are
insignificant in cosmology since their relative contribution to the
energy density of the Universe decreases rapidly with time \cite{textures}.

Even if the nontrivial topology required for the existence
of a defect is absent in a field theory, it may still be possible to have
defect-like solutions. Defects may be {\sl embedded}
in such topologically trivial field theories \cite{embedded}. While
stability of topological defects is guaranteed by topology, embedded
defects are in general unstable under small perturbations.

\subsection{Cosmic Microwave Background Temperature Anisotropies}

The CMB temperature anisotropies offer
a powerful test for theoretical models aiming at describing the early
Universe.  The characteristics of the CMB multipole moments can be
used to discriminate among theoretical models and to constrain
the parameters space.

The spherical harmonic expansion of the CMB temperature anisotropies,
as a function of angular position, is given by
\begin{equation}
\label{dTT}
\frac{\delta T}{T}({\bf n})=\sum _{\ell m}a_{\ell m} {\cal W}_\ell
Y_{\ell m}({\bf n})~\,
\ \ \ \mbox {with}\ \ \ 
a_{\ell m}=\int {\rm
d}\Omega _{{\bf n}}\frac{\delta T}{T}({\bf n})Y_{\ell m}^*({\bf n})~;
\end{equation}
${\cal W}_\ell $ stands for the $\ell$-dependent window function of
the particular experiment.  The angular power spectrum of CMB
temperature anisotropies is expressed in terms of the dimensionless
coefficients $C_\ell$, which appear in the expansion of the angular
correlation function in terms of the Legendre polynomials $P_\ell$:
\begin{equation}
\biggl \langle 0\biggl |\frac{\delta T}{T}({\bf n})\frac{\delta T}{
T}({\bf n}') \biggr |0\biggr\rangle \left|_{{~}_{\!\!({\bf n\cdot
n}'=\cos\vartheta)}}\right. = \frac{1}{4\pi}\sum_\ell(2\ell+1)C_\ell
P_\ell(\cos\vartheta) {\cal W}_\ell^2 ~.
\label{dtovertvs}
\end{equation}
It compares points in the sky separated by an angle $\vartheta$.
Here, the brackets denote spatial average, or expectation values if
perturbations are quantised. Equation (\ref{dtovertvs}) holds only if
the initial state for cosmological perturbations of quantum-mechanical
origin is the vacuum \cite{jm1,jm2}.  The value of $C_\ell$ is
determined by fluctuations on angular scales of the order of
$\pi/\ell$. The angular power spectrum of anisotropies observed today
is usually given by the power per logarithmic interval in $\ell$,
plotting $\ell(\ell+1)C_\ell$ versus $\ell$.

The predictions of the defects models regarding the characteristics of
the CMB spectrum are:
\begin{itemize}
\item
Global ${\cal O}(4)$ textures lead to
a position of the first acoustic peak  at $\ell\simeq 350$
with an amplitude $\sim 1.5$ times higher than the Sachs-Wolfe
plateau \cite{rm}.
\item
Global ${\cal O}(N)$ textures in the
large $N$ limit lead to a quite flat spectrum, with a slow decay after
$\ell \sim 100$ \cite{dkm}. Similar are the predictions of other
global ${\cal O}(N)$ defects \cite{clstrings,num}.
\item
Local cosmic strings predictions are not very well established and
range from an almost flat spectrum \cite{acdkss} to a single wide bump
at $\ell \sim 500$ \cite{mark} with an extremely rapidly decaying
tail.

\end{itemize}
The position and amplitude of the acoustic peaks, as found by the CMB
measurements \cite{maxi,boom,dasi,wmap}, are in disagreement with the
predictions of topological defects models. Thus, CMB measurements rule
out pure topological defects models as the origin of initial density
perturbations leading to the observed structure formation. At this
point one has to ask which are the implications for the high energy
physics models upon which our cosmological model was built. I will
thus first ask whether cosmic strings formation is indeed generic.  I will
address this question in the framework of SUSY GUTs.  I am only
interested in cosmic strings, since I consider gauge theories, for
which domain walls and monopoles are dangerous, while textures are
cosmologically uninteresting \cite{textures}.

\subsection{Genericity of Cosmic Strings Formation within SUSY GUTs}

I will address the question of whether cosmic strings formation is
generic, in the context of SUSY GUTs.  Even though the Standard Model
(SM) has been tested to a very high precision, it is incapable of
explaining neutrino masses \cite{SK,SNO,kamland}.  An extension of the
SM gauge group can be realised within Supersymmetry (SUSY). SUSY
offers a solution to the gauge hierarchy problem, while in the
supersymmetric standard model the gauge coupling constants of the
strong, weak and electromagnetic interactions meet at a single point
$M_{\rm GUT} \simeq (2-3) \times 10^{16}$ GeV.  In addition, SUSY GUTs
can provide the scalar field which could drive inflation, explain the
matter-antimatter asymmetry of the Universe, and propose a candidate,
the lightest superparticle, for cold dark matter.

Within SUSY GUTs there is a large number of SSB patterns leading from
a large gauge group G to the SM gauge group G$_{\rm SM}\equiv$
SU(3)$_{\rm C}\times$ SU(2)$_{\rm L}\times$ U(1)$_{\rm Y}$. The study
of the homotopy group of the false vacuum for each SSB scheme will
determine whether there is defect formation and it will identify the
type of the formed defect. Clearly, if there is formation of domain
walls or monopoles, one will have to place an era of supersymmetric
hybrid inflation to dilute them. To consider a SSB scheme as a
successful one, it should be able to explain the matter/anti-matter
asymmetry of the Universe and to account for the proton lifetime
measurements \cite{SK}.  In what follows, I consider a mechanism of
baryogenesis via leptogenesis, which can be thermal or nonthermal one.
In the case of nonthermal leptogenesis, U(1)$_{\rm B-L}$ (B and L, are
the baryon and lepton numbers, respectively) is a sub-group of the GUT
gauge group, G$_{\rm GUT}$, and B-L is broken at the end or after
inflation. In the case of thermal leptogenesis, B-L is broken
independently of inflation. If leptogenesis is thermal and B-L is
broken before the inflationary era, then one should check whether the
temperature at which B-L is broken, which will define the mass of the
right-handed neutrinos, is smaller than the reheating temperature
which should be lower than the limit imposed by the gravitino. To
ensure the stability of proton, the discrete symmetry Z$_2$, which is
contained in U(1)$_{\rm B-L}$, must be kept unbroken down to low
energies. This implies that the successful SSB schemes should end at
G$_{\rm SM}\times$ Z$_2$.  I will then examine how often cosmic
strings have survived at the inflationary era, within all acceptable
SSB patterns.

To accomplish this task one has to choose the large gauge group
G$_{\rm GUT}$.  In Ref. \cite{jrs} this study has been done explicitly
for a large number of simple Lie groups. Since I consider GUTs based
on simple gauge groups, the type of supersymmetric hybrid inflation
will be of the F-type. The minimum rank of G$_{\rm GUT}$ has to be at
least equal to 4, to contain the G$_{\rm SM}$ as a subgroup.  Then one
has to study the possible embeddings of G$_{\rm SM}$ in G$_{\rm GUT}$
to be in agreement with the SM phenomenology and especially with the
hypercharges of the known particles. Moreover, the group must include
a complex representation, needed to describe the SM fermions, and it
must be anomaly free.  Since, in principle, ${\rm SU}(n)$ may not be
anomaly free, I assume that the ${\rm SU}(n)$ groups which I use have
indeed a fermionic representation that certifies that the model is
anomaly free. I set as the upper bound on the rank $r$ of the group,
$r\leq 8$. Clearly, the choice of the maximum rank is in principle
arbitrary.  This choice could, in a sense, be motivated by the
Horava-Witten \cite{hw} model, based on ${\rm E}_8\times {\rm
E}_8$. Thus, the large gauge group G$_{\rm GUT}$ could be one of the
following: SO(10), E$_6$, SO(14), SU(8), SU(9); flipped SU(5) and
[SU(3)]$^3$ are included within this list as subgroups of SO(10) and
E$_6$, respectively.

A detailed study of all the SSB schemes which bring us from G$_{\rm
GUT}$ down to the SM gauge group G$_{\rm SM}$, by one or more
intermediate steps, shows that cosmic strings are generically formed
at the end of hybrid inflation.  If the large gauge group G$_{\rm
GUT}$ is SO(10) then cosmic strings formation is unavoidable
\cite{jrs}.  For ${\rm E}_6$ it depends whether one considers thermal
or nonthermal leptogenesis. More precisely, under the assumption of
nonthermal leptogenesis then cosmic strings formation is
unavoidable. If I consider thermal leptogenesis then cosmic strings
formation at the end of hybrid inflation arises in $98\%$ of the
acceptable SSB schemes \cite{jm}.  If the requirement of having Z$_2$
unbroken down to low energies is relaxed and thermal leptogenesis is
considered as being the mechanism for baryogenesis, then cosmic
strings formation accompanies hybrid inflation in $80\%$ of the SSB
schemes ~\cite{jm}.  The SSB schemes of SU(6) and SU(7) down to the
G$_{\rm SM}$ which could accommodate an inflationary era with no
defect (of any kind) at later times are inconsistent with proton
lifetime measurements and minimal SU(6) and SU(7) do not predict
neutrino masses \cite{jrs}, implying that these models are
incompatible with high energy physics phenomenology. Higher
rank groups, namely SO(14), SU(8) and SU(9), should in general lead to
cosmic strings formation at the end of hybrid inflation. In all these
schemes, cosmic strings formation is sometimes accompanied by the
formation of embedded strings. The strings which form at the end of
hybrid inflation have a mass which is proportional to the inflationary
scale.

\subsection{Mixed Models}
Since cosmic strings are expected to be generically formed in the
context of SUSY GUTs, one should consider {\sl mixed perturbation
models} where the dominant r\^ole is played by the inflaton field but cosmic
strings have also a contribution, small but not negligible.  
Restricting ourselves to the angular power spectrum, we can remain in the
linear regime. In this case, 
\begin{equation}
C_\ell =   \alpha     C^{\scriptscriptstyle{\rm I}}_\ell
         + (1-\alpha) C^{\scriptscriptstyle{\rm S}}_\ell~,
\label{cl}
\end{equation}
where $C^{\scriptscriptstyle{\rm I}}_\ell$ and $C^{\scriptscriptstyle
{\rm S}}_\ell$ denote the (COBE normalized) Legendre coefficients due
to adiabatic inflaton fluctuations and those stemming from the cosmic
strings network, respectively. The coefficient $\alpha$ in
Eq.~(\ref{cl}) is a free parameter giving the relative amplitude for
the two contributions.  Comparing the $C_\ell$, given by
Eq.~(\ref{cl}), with data obtained from the most recent CMB
measurements, one gets that a cosmic strings contribution to the
primordial fluctuations higher than $14\%$ is excluded up to $95\%$
confidence level \cite{bprs,pogosian,wyman}.  In what follows, I will
be on the conservative side and and I will not allow cosmic strings to
contribute more than $10\%$ to the CMB temperature anisotropies.
 
\subsection{Supersymmetric Hybrid Inflation}

Inflation remains the most appealing scenario for describing the early
Universe.  Inflation essentially consists of a phase of accelerated
expansion which took place at a very high energy scale. However,
despite its success, it faces a number of questions, as for example
how generic is the onset of inflation \cite{ecms} and how one can
guarantee a natural and successful inflationary model. Unfortunately,
inflation is still a paradigm in search of a model. I will discuss
two well-studied inflationary models in the framework of
supersymmetry, namely F/D-term inflation. 

\subsubsection{F-term Inflation}

F-term inflation can be naturally accommodated in the framework of
GUTs when a GUT gauge group G$_{\rm GUT}$ is broken down to the
G$_{\rm SM}$ at an energy $M_{\rm GUT}$ according to the scheme
\begin{equation}
{\rm G}_{\rm GUT} \stackrel{M_{\rm GUT}}{\hbox to 0.8cm
{\rightarrowfill}} {\rm H}_1 \xrightarrow{9}{M_{\rm
infl}}{1}{\Phi_+\Phi_-} {\rm H}_2 {\longrightarrow} {\rm G}_{\rm SM}~;
\end{equation}
$\Phi_+, \Phi_-$ is a pair of GUT Higgs superfields in nontrivial
complex conjugate representations, which lower the rank of the group
by one unit when acquiring nonzero vacuum expectation value. The
inflationary phase takes place at the beginning of the symmetry
breaking ${\rm H}_1\stackrel{M_{\rm infl}}{\longrightarrow} {\rm
H}_2$.

F-term inflation is based on the globally supersymmetric
renormalisable superpotential
\begin{equation}\label{superpot}
W_{\rm infl}^{\rm F}=\kappa  S(\Phi_+\Phi_- - M^2)~,
\end{equation}
where $S$ is a GUT gauge singlet left handed superfield, $\Phi_+$ and
$\Phi_-$ are defined above; $\kappa$ and $M$ are two constants ($M$
has dimensions of mass) which can be taken positive with field
redefinition.  The chiral superfields $S, \Phi_+, \Phi_-$ are
taken to have canonical kinetic terms.  This superpotential is the
most general one consistent with an R-symmetry under which $W
\rightarrow e^{i \beta} W~, \Phi_- \rightarrow e^{-i \beta} \Phi_-~,
\Phi_+ \rightarrow e^{i \beta} \Phi_+$, and $ S \rightarrow e^{i
\beta} S$. An R-symmetry can ensure that the rest of the
renormalisable terms are either absent or irrelevant.

The scalar potential reads
\begin{equation}
\label{scalpot1}
V(\phi_+,\phi_-, S)= |F_{\Phi_+}|^2+|F_{\Phi_-}|^2+|F_ S|^2
+\frac{1}{2}\sum_a g_a^2 D_a^2~.
\end{equation}
The F-term is such that $F_{\Phi_i} \equiv |\partial W/\partial
\Phi_i|_{\theta=0}$, where we take the scalar component of the
superfields once we differentiate with respect to $\Phi_i=\Phi_+,
\Phi_-,  S$. The D-terms are
\begin{equation}
D_a=\bar{\phi}_i\,{(T_a)^i}_j\,\phi^j +\xi_a~,
\end{equation}
with $a$ the label of the gauge group generators $T_a$, $g_a$ the
gauge coupling, and $\xi_a$ the Fayet-Iliopoulos term. By definition,
in the F-term inflation the real constant $\xi_a$ is zero; it can only
be nonzero if $T_a$ generates an extra U(1) group.  In the context of
F-term hybrid inflation, the F-terms give rise to the inflationary
potential energy density, while the D-terms are flat along the
inflationary trajectory, thus one may neglect them during inflation.

The potential has one valley of local minima, $V=\kappa^2 M^4$, for
$S> M $ with $\phi_+ = \phi_-=0$, and one global supersymmetric
minimum, $V=0$, at $S=0$ and $\phi_+ = \phi_- = M$. Imposing initially
$ S \gg M$, the fields quickly settle down the valley of local
minima.  Since in the slow roll inflationary valley, the ground state
of the scalar potential is nonzero, SUSY is broken.  In the tree
level, along the inflationary valley the potential is constant,
therefore perfectly flat. A slope along the potential can be generated
by including the one-loop radiative corrections. Thus, the scalar
potential gets a little tilt which helps the inflaton field $S$ to
slowly roll down the valley of minima. The one-loop radiative
corrections to the scalar potential along the inflationary valley,
lead to an effective potential
\cite{DvaShaScha,Lazarides,SenoSha,rs1}
\begin{equation}
\label{VexactF}
V_{\rm eff}^{\rm F}(|S|)=\kappa^2M^4\big\{1+\frac{\kappa^2
\cal{N}}{32\pi^2}\big[2\ln\frac{|S|^2\kappa^2}{\Lambda^2}
+\big(\frac{|S|^2}{ M^2}+1\big)^2\ln\big(1+\frac{M^2}{|S|^2})
+\big(\frac{|S|^2}{M^2}-1\big)^2\ln\big(1-\frac{M^2}{|S|^2}\big)\big]\big\}
~;
\end{equation}
$\Lambda$ is a renormalisation scale 
and $\cal{N}$ stands for the dimensionality
of the representation to which the complex scalar components $\phi_+,
\phi_-$ of the chiral superfields $\Phi_+, \Phi_-$ belong. 

Considering only large angular scales, one can get the contributions
to the CMB temperature anisotropies analytically. In
Ref. \cite{rs1}, the Sachs-Wolfe effect has been explicitly
calculated. The quadrupole anisotropy has one contribution coming from
the inflaton field, calculated using Eq.~(\ref{VexactF}), and one
contribution coming from the cosmic strings network, given by
numerical simulations \cite{ls}. Fixing the number of e-foldings to
60, then for a given gauge group G$_{\rm GUT}$, the inflaton and cosmic
strings contribution to the CMB depend on the superpotential coupling
$\kappa$, or equivalently on the symmetry breaking scale $M$
associated with the inflaton mass scale, which coincides with the
string mass scale. The total quadrupole anisotropy has to be
normalised to the COBE data.  In Ref .~\cite{rs1} we have found
that the cosmic strings contribution is consistent with the CMB
measurements, provided
\begin{equation}
M\lsim 2\times 10^{15} {\rm GeV} ~~\Leftrightarrow ~~\kappa \lsim
7\times10^{-7}~.
\end{equation}
This constraint on $\kappa$ is in agreement with the one found in 
Ref.~\cite{lk}. 
Strictly speaking the above condition was found in the context of SO(10) 
gauge group, but the conditions imposed in the context of other gauge
groups are of the same order of magnitude since $M$ is a slowly varying
function of the dimensionality ${\cal N}$ of the representations to which 
the scalar components of the chiral Higgs superfields belong.

The superpotential coupling $\kappa$ is also subject to the gravitino
constraint which imposes an upper limit to the reheating temperature,
to avoid gravitino overproduction. Within the framework of SUSY GUTs
and assuming a see-saw mechanism to give rise to massive neutrinos,
the inflaton field decays during reheating into pairs of right-handed
neutrinos.  This constraint on the reheating temperature can be
converted to a constraint on the parameter $\kappa$. The gravitino
constraint on $\kappa$ reads \cite{rs1} $\kappa \lsim 8\times
10^{-3}$, which is a weaker constraint.

Concluding, F-term inflation leads generically to cosmic strings
formation at the end of the inflationary era. The cosmic strings
formed are of the GUT scale. This class of models can be compatible
with CMB measurements, provided the superpotential coupling is smaller
than $10^{-6}$.  This tuning of the free parameter $\kappa$ can be
softened if one allows for the curvaton mechanism.  According to the
curvaton mechanism \cite{lw2002,mt2001}, another scalar field, called
the curvaton, could generate the initial density perturbations whereas
the inflaton field is only responsible for the dynamics of the
Universe. The curvaton is a scalar field, that is sub-dominant during
the inflationary era as well as at the beginning of the radiation
dominated era which follows the inflationary phase. There is no
correlation between the primordial fluctuations of the inflaton and
curvaton fields. Clearly, within supersymmetric theories such scalar
fields are expected to exist. In addition, embedded strings, if they
accompany the formation of cosmic strings, they may offer a natural
curvaton candidate, provided the decay product of embedded strings
gives rise to a scalar field before the onset of inflation.
Considering the curvaton scenario, the coupling $\kappa$ is only
constrained by the gravitino limit. More precisely, assuming the
existence of a curvaton field, there is an additional contribution to
the temperature anisotropies. The WMAP CMB measurements impose
\cite{rs1} the following limit on the initial value of the curvaton
field
\begin{equation}
{\cal\psi}_{\rm init} \lsim 5\times 10^{13}\,\left( 
\frac{\kappa}{10^{-2}}\right){\rm GeV}~,
\end{equation}
provided the parameter $\kappa$ is in the range $[10^{-6},~1]$.

\subsubsection{D-term Inflation}
D-term inflation received a lot of interest, mainly because it is not
plagued by the {\sl Hubble-induced mass} problem, and in addition it
can be easily implemented in string theory. D-term inflation is
derived from the superpotential
\begin{equation}
\label{superpotD}
W^{\rm D}_{\rm infl}=\lambda S \Phi_+\Phi_-~;
\end{equation}
$S, \Phi_-, \Phi_+$ are three chiral superfields and $\lambda$ is the
superpotential coupling.  D-term inflation requires the existence of a
nonzero Fayet-Iliopoulos term $\xi$, which can be added to the
Lagrangian only in the presence of an extra U(1) gauge symmetry, under
which, the three chiral superfields have charges $Q_S=0$,
$Q_{\Phi_+}=+1$ and $Q_{\Phi_-}=-1$, respectively. This extra U(1)
gauge symmetry symmetry can be of a different origin; hereafter we
consider a nonanomalous U(1) gauge symmetry.  Thus, D-term inflation
requires a scheme, like
\begin{equation}
{\rm G}_{\rm GUT}\times {\rm U}(1) \stackrel{M_{\rm GUT}}{\hbox to
0.8cm{\rightarrowfill}} {\rm H} \times {\rm U}(1) \xrightarrow{9}{M_{\rm
nfl}}{1}{\Phi_+\Phi_-} {\rm H} \rightarrow {\rm G}_{\rm SM}~.
\end{equation}
The symmetry breaking at the end of the inflationary phase implies
that cosmic strings are always formed at the end of D-term hybrid
inflation. To avoid cosmic strings, several mechanisms have been
proposed which either consider more complicated models or require
additional ingredients. For example, one can add a nonrenormalisable
term in the potential \cite{shifted}, or add an additional discrete
symmetry \cite{smooth}, or consider GUT models based on nonsimple
groups \cite{mcg}, or introduce a new pair of charged superfields
\cite{jaa} so that cosmic strings formation is avoided within D-term
inflation. In what follows, I will show that standard D-term inflation
leading to cosmic strings production is still compatible with CMB data
since cosmic strings contribution to the CMB data is not constant nor
dominant. This implies that one does not have to invoke some new
physics. The reader can find a detailed study in Ref.~\cite{rs1,rs2}.

In the global supersymmetric limit,
Eqs.~(\ref{scalpot1}),~(\ref{superpotD}) lead to the following
expression for the scalar potential
\begin{equation}
\label{VtotD}
V^{\rm D}(\phi_+,\phi_-,S) = \lambda^2
\left[\,|S|^2(|\phi_+|^2+|\phi_-|^2) + |\phi_+\phi_-|^2 \right]
+\frac{g^2}{2}(|\phi_+|^2-|\phi_-|^2+\xi)^2~,
\end{equation}
where $g$ is the gauge coupling of the U(1) symmetry and $\xi$ is a
Fayet-Iliopoulos term, chosen to be positive.

In D-term inflation, as opposed to F-term inflation, the inflaton mass
acquires values of the order of Planck mass, and therefore, the
correct analysis must be done in the framework of SUGRA. The SSB of
SUSY in the inflationary valley introduces a splitting in the masses
of the components of the chiral superfields $\Phi_\pm$. As a result,
we obtain \cite{rs2} two scalars with squared masses
$m^2_{\pm}=\lambda^2|S|^2 \exp\left(|S|^2/M^2_{\rm Pl}\right)\pm g^2
\xi$ and a Dirac fermion with squared mass $m_{\rm f}^2=\lambda^2|S|^2
\exp\left(|S|^2/M^2_{\rm Pl}\right)$.  Thus, calculating the radiative
corrections, the effective scalar potential for minimal supergravity
reads~\cite{rs1,rs2}
\begin{eqnarray}
\label{vDsugra}
V_{\rm eff} 
&=&\frac{g^2\xi^2}{2}\big\{1+\frac{g^2}{16 \pi^2}
\times\big[2\ln\frac{|S|^2\lambda^2}{\Lambda^2}e^{\frac{|S|^2}{
M^2_{\rm Pl}}}
+\big( \frac{\lambda^2 |S|^2}{
g^2\xi}e^{\frac{|S|^2}{M_{\rm Pl}^2}} +1\big)^2
\ln\big(1+\frac{ g^2\xi}{\lambda^2 |S|^2 }e^{-\frac{|S|^2}{ M_{\rm Pl}^2}}
\big)\nonumber\\
&&
+\big( \frac{\lambda^2 |S|^2}{
g^2\xi}e^{\frac{|S|^2}{ M_{\rm Pl}^2}} -1\big)^2
\ln\big(1-\frac{ g^2\xi}{\lambda^2 |S|^2 }e^{-\frac{|S|^2}{ M_{\rm Pl}^2}}
\big)\big]\big\}
\end{eqnarray}
In Refs.~\cite{rs1,rs2}, we have properly addressed the question
of cosmic strings contribution to the CMB data and we got that
standard D-term inflation can be compatible with measurements; the
cosmic strings contribution to the CMB is actually
model-dependent. Our most important finding was that cosmic strings
contribution is not constant, nor is it always dominant.
   
More precisely, we obtained \cite{rs1,rs2} that $g\gsim 2\times
 10^{-2}$ is incompatible with the allowed cosmic strings contribution
 to the WMAP measurements.  For $g\lsim 2\times 10^{-2}$, the
 constraint on the superpotential coupling $\lambda$ reads $\lambda
 \lsim 3\times 10^{-5}$.  SUGRA corrections impose in addition a lower
 limit to $\lambda$.  The constraints induced on the couplings by the
 CMB measurements can be expressed \cite{rs1,rs2} as a single
 constraint on the Fayet-Iliopoulos term $\xi$, namely $\sqrt\xi \lsim
 2\times 10^{15}~{\rm GeV}$.

Concluding, standard D-term inflation always leads to cosmic strings
formation at the end of the inflationary era. The cosmic strings
formed are of the GUT scale. This class of models is still compatible
with CMB measurements, provided the couplings are small enough.  As in
the case of F-term inflation the fine tuning of the couplings can be
softened provided one considers the curvaton mechanism. In this case,
the imposed CMB constraint on the initial value of the curvaton field
reads \cite{rs1,rs2}
\begin{equation}
\psi_{\rm init}\lsim 3\times 10^{14}\left(\frac{g}{ 10^{-2}}\right) ~{\rm GeV},
\end{equation}
for $\lambda\in [10^{-1}, 10^{-4}]$.

Our conclusions are still valid in the revised version of D-term
inflation, in the framework of SUGRA with constant Fayet-Iliopoulos
terms. In the context of N=1, 3+1 SUGRA, the presence of constant
Fayet-Iliopoulos terms shows up in covariant derivatives of all
fermions. In addition, since the relevant local U(1) symmetry is a
gauged R-symmetry \cite{toine2}, the constant Fayet-Iliopoulos terms
also show up in the supersymmetry transformation laws.  In
Ref.~\cite{toine1} there were presented all corrections of order
$g\xi/M_{\rm Pl}^2$ to the classical SUGRA action required by local
supersymmetry.  Under U(1) gauge transformations in the directions in
which there are constant Fayet-Iliopoulos terms $\xi$, the
superpotential must transform as \cite{toine2}
\begin{equation}
\delta W=-i\frac{g\xi}{ M_{\rm Pl}^2}W~,
\end{equation}
otherwise the constant Fayet-Iliopoulos term $\xi$ vanishes. This
requirement is consistent with the fact that in the gauge theory at
$M_{\rm Pl}\rightarrow \infty$ the potential is U(1) invariant. To
promote the simple SUSY D-term inflation model, Eq.~(\ref{superpotD}),
to SUGRA with constant Fayet-Iliopoulos terms, one has to change the
charge assignments for the chiral superfields, so that the
superpotential transforms under local R-symmetry \cite{toine1}. In
SUSY, the D-term potential is neutral under U(1) symmetry, while in
SUGRA the total charge of $\Phi_\pm$ fields does not vanish but is
equal to $-\xi/M_{\rm Pl}^2$. More precisely, the D-term contribution
to the scalar potential $V$ [see Eq.~(\ref{VtotD})], should be
replaced by $(g^2/2)(q_+|\phi_+|^2+q_-|\phi_-|^2+\xi)^2$ where
\begin{equation}
q_\pm=\pm 1-\rho_\pm\frac{\xi}{ M_{\rm Pl}^2}\ \ \ \ \mbox {with} \ \ \
\ \rho_++\rho_-=1~.
\end{equation}
In addition, the squared masses of the scalar components $\phi_\pm$
become 
\begin{equation}
m^2_{\pm}=\lambda^2|S|^2 \exp\left(|S|^2/M^2_{\rm Pl}\right)\pm g^2
\xi q_\pm~;
\end{equation}
the Dirac fermion mass remains unchanged.  However, sine for the
limits we imposed on the Fayet-Iliopoulos term $\xi$, the correction
$\xi/M_{\rm Pl}^2$ is $\sim 10^{-6}$, I conclude that our results also
also valid in the revised version of D-term inflation within SUGRA.

\section{Superstrings as Cosmic Strings Candidates}

In the context of perturbative string theory, superstrings of cosmic
size were excluded, mainly because they should have too large a
tension. More precisely, perturbative strings have a tension close to
the Planck scale, producing CMB inhomogeneities far larger than
observed. Moreover, since the scale of their tension also exceeds the
upper bound on the energy scale of the inflationary vacuum, such
strings could have only been produced before inflation, and therefore
diluted. In addition, there are instabilities that would prevent such long
strings from surviving on cosmic time scales \cite{witten}.
Thus, for years was a clear distinction between
fundamental strings and cosmic strings.

Recently, this whole picture has changed. In addition to the
fundamental F-strings, there are also D-strings as a particular case
of higher-dimensional D$p$-branes (D stands for Dirichlet and $p$
denotes the dimensionality of the brane), partially wrapped on compact
cycles resulting to only one noncompact dimension.  In the braneworld
approach, our  Universe represents a D3-brane on which open
strings can end \cite{Pol}, embedded in a higher dimensional space,
called the bulk. Brane interactions can unwind and evaporate higher
dimensional D$p$-branes so that we are left with D3-branes embedded in
a higher dimensional bulk; one of these D3-branes plays the r\^ole of
our Universe \cite{dks}.  Since gauge charges are attached to the ends
of strings, gauge particles and fermions can propagate only along the
D3-branes while gravitons (and dilatons, ...)  which are closed string
modes can move in the bulk. Since gravity has been probed only down to
scales of about $ 0.1$mm, the dimensions of the bulk can be much
larger than the string scale. In the braneworld context, the extra
dimensions can even be infinite, if the geometry is nontrivial 
\cite{RSII}. Large extra dimensions can be employed to
address the hierarchy problem \cite{Ark}, a result which lead to an
increasing interest in braneworld scenarios.  Apart from the D$p$-branes,
there are also antibranes, $\bar Dp$-branes, which differ from the
D$p$-branes by having an equal and opposite conserved Ramond-Ramond
charge, which implies an attractive force between them.

Braneworld cosmology can also offer a natural inflationary scenario.
Assuming the early Universe contained an extra brane and antibrane,
then an inflationary era could be driven by the potential between the
two branes, while the separation between the branes will play the
r\^ole of the inflaton. The inflaton potential is rather flat when the
branes are separated and steepens as they approach, until at some
point a field becomes tachyonic, which indicates an instability,
leading to a rapid brane-antibrane annihilation.

  D-brane-antibrane inflation leads to the abundant production of
lower dimensional D-branes that are one-dimensional in the noncompact
directions \cite{dstr}.  Luckily, zero-dimensional defects (monopoles)
and two-dimensional ones (domain walls), which are cosmologically
undesirable, are not produced. In these models, the large compact
dimensions and the large warp factors allow cosmic superstring
tensions to be in the range between $10^{-11}< G\mu < 10^{-6}$,
depending on the model.

Cosmic superstrings share a number of properties with the cosmic
strings, but there are also differences which may lead to distinctive
observational signatures. String intersections lead to
intercommutation and loop production. For cosmic strings the
probability of intercommutation ${\cal P}$ is equal to 1, whereas this
is not the case for F- and D-strings. Clearly, D-strings can miss each
other in the compact dimension, leading to a smaller ${\cal P}$, while
for F-strings the scattering has to be calculated quantum mechanically
since these are quantum mechanical objects.

The collisions between all possible pairs of superstrings have been
studied in string perturbation theory \cite{jjp}. For F-strings, the
reconnection probability is of the order of $g_{\rm s}^2$, where
$g_{\rm s}$ stands for the string coupling.  For F-F string
collisions, it was found \cite{jjp} that the reconnection probability
$\cal P$ is $10^{-3}\lsim {\cal P}\lsim 1$. For D-D string collisions,
one has $10^{-1}\lsim{\cal P}\lsim 1$. Finally, for F-D string
collisions, the reconnection probability can take any value between 0
and 1.  These results have been confirmed \cite{hh1} by a quantum
calculation of the reconnection probability for colliding D-strings.
Similarly, the string self-intersection probability is
reduced. Moreover, when D- and F-strings meet they can form a
three-string junction, with a composite DF-string. It is also possible
in IIB string theory to have bound $(p,q)$ states of $p$ F-strings and
$q$ D-strings, where $p$ and $q$ are coprime. This leads to the
question of whether there are frozen networks dominating the matter
content of the Universe, or whether scaling solutions can be achieved.

To study the evolution of cosmic superstrings, I have performed
numerical simulations \cite{ms} of independent stochastic networks of
D- and F-strings. I found that the characteristic length scale $\xi$,
which gives the typical distance between the nearest string segments
and the typical curvature of strings, grows linearly with time
\begin{equation}
 \xi(t)\propto \zeta t ~,
\end{equation}
where the slope $\zeta$ depends on the reconnection probability ${\cal
P}$, and on the energy of the smallest allowed loops (i.e., the energy
cutoff).  For reconnection (or intercommuting) probability in the
range $10^{-3}\lsim {\cal P} \lsim 0.3$, I found \cite{ms}
\begin{equation}
\zeta \propto \sqrt{\cal P} \Rightarrow \xi(t)\propto \sqrt{\cal P} t~,
\label{law}
\end{equation}
in agreement with my old results \cite{sv}.  I thus disagree with the
statement that $\xi(t)\propto {\cal P} t$.  In Ref.~\cite{jst} it is
claimed that the energy density of longs strings $\rho_{\rm l}$
evolves as $\dot\rho_{\rm l}=2(\dot a/a)\rho_{\rm l}-{\cal
P}(\rho_{\rm l}/\xi)$, where $H=\dot a/a$ is the Hubble constant.
Then substituting the ansatz $\xi(t)=\gamma(t)t$, the authors of
Ref.~\cite{jst} obtain $\dot\gamma=-[1/(2t)](\gamma-{\cal P})$, during
the radiation dominated era. Since this equation has a stable fixed
point at $\gamma(t)={\cal P}$, the authors state \cite{jst} that
$\xi\simeq {\cal P} t$. My disagreement with Ref.~\cite{jst} is based
on the fact that intersections between two long strings is not the
most efficient mechanism for energy loss of the string network. The
possible string intersections can be divided into three possible
cases: (i) two long strings collide in one point and exchange partners
with intercommuting probability ${\cal P}_1$; (ii) two strings collide
in two points and exchange partners chopping off a small loop with
intercommuting probability ${\cal P}_1^2$; and (iii) one long string
self-intersects in one point and chops off a loop with intercommuting
probability ${\cal P}_2$, which in general is different than ${\cal
P}_1$.  Clearly, only cases (ii) and (iii) lead to a closed loop
formation and therefore remove energy from the long string network.
Between cases (ii) and (iii), only case (iii) is an efficient way of
forming loops and therefore dissipating energy.  I have checked
numerically \cite{ms} that case (iii) appears more often than case
(ii), and besides, case (ii) has in general a smaller probability,
since one expects that ${\cal P}_1\sim {\cal P}_2$.  However, the
heuristic argument employed in Ref.\cite{jst} does not refer to
self-string intersections (i.e, case (iii)); it only applies to
intersections between two long strings. This is clear since
intersections between two long strings depend on the string velocity,
however self-string intersections should not depend on how fast the
string moves. In other words, a string can intersect itself even if it
does not move but it just oscillates locally.

Studying the time evolution of the slope $\zeta$, I found \cite{ms}
that it reaches a constant value at relatively the same time $t$ for
various values of ${\cal P}$, which implies that the long strings
reach scaling. This result has been confirmed by studying numerically
the behavior of a network of interacting Dirichlet-fundamental strings
$(p,q)$ in Ref.~\cite{ep}. To model $(p,q)$ strings arising from
compactifications of type IIB string theory, the authors studied
\cite{ep} the evolution of nonabelian string networks. The positive
element of such nonabelian networks is that they contain multiple
vertices where many different types of string join together.  Such
networks have the potential of leading to a string dominated Universe
due to tangled networks of interacting $(p,q)$ strings that freeze.
It was shown \cite{ep} that such freezing does not take place and the
network reaches a scaling limit. In this field theory approach however
strings are not allowed to have different tensions, which is a
characteristic property of cosmic superstrings. Recently, this has
been done in the context of modelling $(p,q)$ cosmic superstrings
\cite{tww}. It was found that such networks rapidly approach a stable
scaling solution, where once scaling is reached, only a small number
of the lowest tension states is populated substantially. An
interesting question is to find out whether the field theory approach
of Ref.~\cite{ep} mimics the results of the modelling approach of
Ref.~\cite{tww}.

The cosmic superstring network is characterised \cite{ms} by two
components: there are a few long strings with a scale-invariant
evolution; the characteristic curvature radius of long strings, as
well as the typical separation between two long strings are both
comparable to the horizon size, $\xi(t)\simeq {\sqrt {\cal P}} t$, and there
is a large number of small closed loops having sizes $\ll t$.
Assuming there are string interactions, the long strings network will
reach an asymptotic energy density where
\begin{equation}
\rho_{\rm l}=\frac{\mu}{{\cal P} t^2}~.
\end{equation}
Thus, for fixed linear mass density, the cosmic superstring energy
density may be higher than the field theory case, but at most only by
one order of magnitude. More precisely, the fraction of the total
density in the form of strings in the radiation-dominated era reads
\begin{equation}
\frac{\rho_{\rm str}}{\rho_{\rm total}}=\frac{32\pi}{ 3} \frac{G\mu}{{\cal P}}~.
\end{equation}

Oscillating string loops loose energy by emitting graviton, dilaton
and Ramond-Ramond (RR) fields. Accelerated cosmic strings are sources
of gravitational radiation, in particular from the vicinity of the
cusps where the string velocity approaches the speed of
light. Similarly, cosmic superstrings emit gravity waves but since the
intercommutation probability is less than unity, their network is
denser with more cusps, resulting in an enhancement of the emitted
gravitational radiation. As it was pointed out \cite{dv}, the
gravitational wave bursts emitted from cusps of oscillating string or
superstring loops could be detectable with the gravitational-wave
interferometers LIGO/VIRGO and LISA.

One can place constraints on the energy scale of cosmic strings from
the observational bounds on dilaton decays \cite{tdv}.  Considering
that the dilaton lifetime is in the range $10^7{\rm s}\lsim \tau\lsim
t_{\rm dec}$, I obtained an upper bound $\eta\lsim {\cal
P}^{-1/3}\lsim 10^{11}{\rm GeV}$ for the energy scale of cosmic
superstrings, which determines the critical temperature for the
transition leading to string formation. A lower reconnection
probability allows a higher energy scale of strings,  at most by one
order of magnitude.

\section{Cosmic Strings in the Sky}

As a theoretician, I believe that it is of great importance to get
observational support for the existence of cosmic strings.
Unfortunately, up to recently the attempts to find cosmic
strings in the sky were unsuccessful.

A Russian-Italian collaboration claims to have found the
first signature of a cosmic string in the sky. More precisely, the
authors of Refs.~\cite{CSL1a,CSL1b,CSL1c} point out that the peculiar
properties of the gravitational lens CSL-1 (Capodimonte - Sternberg
Lens Candidate no.1) could be only explained as the first case of
lensing by a cosmic string. CSL-1, found in the OACDF (Ossservatorio
Astronomico di Capodimonte - Deep Field) consists of two identical
images, separated by $1.9''$.  The two sources have very similar
morphology, namely they consist or a bright nucleus surrounded by a
faint halo with undistorted and quite circular isophotes. The most
relevant feature of these images is indeed that their isophotes appear
to be undistorted. The performed photometric and spectroscopic
analysis \cite{CSL1a,CSL1c} revealed that both the two components of
CSL-1 are giant elliptical galaxies at redshift $z=0.46$.  The
possibility that CSL-1 could be interpreted as the projection of two
giant elliptical galaxies, identical at a $99\%$ confidence level, has
been disregarded \cite{CSL1a} as unlikely. Moreover, the peculiar
properties of CSL-1 cannot be explained in terms of lensing by a
compact lens model, since a usual gravitational lens created by a
bound clump of matter lead to inhomogeneous gravitational fields which
always distort background extended images. Thus, the most favorite
explanation of CSL-1 in the framework of gravitational lensing is,
according to the authors of Refs.~\cite{CSL1a,CSL1b}, that of lensing
by a cosmic string.  Assuming that CSL-1 is indeed the first lensing by
a cosmic string then the observed separation of the two images
corresponds to a particular value for the deficit angle which implies
that $G\mu>4\times 10^{-7}$; for kinky strings $G\mu$ could be less.
As it was recently pointed out \cite{bsmw}, high string velocities
enhance lensing effects by a factor $1/\sqrt{1-{\bf v}^2}$, where 
${\bf v}$ stands for the string velocity. This decreases
the lower bound on $G\mu$ placed by CSL-1.

In Ref.~\cite{CSL1b}, the authors study the statistics of
lens candidates in the vicinity of CSL-1. They claim that they expect
7-9 lens candidates, which is a relatively high number with respect to
the one expected from normal gravitational lens statistics. This
excess of gravitational lens candidates in the neighborhood of CSL-1 is
claimed \cite{CSL1b} to be compatible with the proposed in
Ref.~\cite{CSL1a} cosmic string scenario. As the authors however state
\cite{CSL1b} only once we have spectroscopic studies of these
candidates, we will be able to extract robust conclusions.
 
It is crucial to confirm or infirm this finding by further and
independent studies.

\section{Conclusions}

In this lecture I presented the story of cosmic strings, as we know it
at present. Cosmic strings are expected to be generically formed in a
large class of models based on SUSY GUTs. If the predictions of cosmic
strings are inconsistent with the various measurements, then either
the theories which predict the formation of cosmic strings are
altogether wrong, or the models have to be more complicated to avoid
stings formation.  Luckily, neither is needed. The free parameters of
the models can be constrained so that there is agreement between
predictions and measurements. Cosmological inflation is an attractive
model with however too many possible choices. It is crucial to find
out which are the natural inflationary models and to constrain their
free parameters. Therefore, the constraints imposed by the
cosmological implications of cosmic strings are indeed important.  The
recent proposal that cosmic superstrings can be considered as cosmic
strings candidates opens new perspectives on the theoretical point of
view. Therefore, even though cosmic strings cannot play a dominant
r\^ole in structure formation, one has to consider them as a
sub-dominant partner of inflation. The possible observational support
which was announced recently is of course a major issue and cosmic
strings have just entered a new flourishing era.

\begin{acknowledgement}
It is a pleasure to thank the organizers, and in particular Mariusz
Dabrowski, for the interesting and stimulating meeting {\sl Pomeranian
Workshop in Fundamental Cosmology}. I would like also to thank my
colleagues with whom I have collaborated through the years in the
various aspects covered in this lecture.
\end{acknowledgement}

\end{document}